\documentclass[%
reprint,
superscriptaddress,
nofootinbib,
amsmath,amssymb,
aps,
pre,
]{revtex4-2}

\usepackage{CJK}%
\usepackage{graphicx}%
\usepackage{dcolumn}%
\usepackage{bm}%
\usepackage{hyperref}
\hypersetup{breaklinks=true,   %
            colorlinks=true,
            linkcolor=blue,
            citecolor=magenta, 
            urlcolor=[rgb]{0.176, 0.184, 0.569}
            }%

\usepackage{newtxtext}
\usepackage{newtxmath}

\makeatletter
\renewcommand{\fnum@figure}{\textbf{Fig.~\thefigure}}
\renewcommand{\@caption@fignum@sep}{\textbf{.}\enspace}
\makeatother

\begin{document}

\begin{CJK*}{UTF8}{gbsn}

\title{Divergence of detachment forces in the finite Voronoi model}%

\author{Wei Wang (汪巍)}
\affiliation{%
Department of Physics and Astronomy, Johns Hopkins University, Baltimore, Maryland 21218, USA\\
}%

\author{Brian A. Camley}
\affiliation{%
Department of Physics and Astronomy, Johns Hopkins University, Baltimore, Maryland 21218, USA\\
}%
\affiliation{%
Department of Biophysics, Johns Hopkins University, Baltimore, Maryland 21218, USA\\
}

\begin{abstract}
Detachment and fracture are central to many tissue-level processes, but they are challenging to simulate with Voronoi-type models that typically assume a confluent tissue.
Here we analyze the finite Voronoi model, a nonconfluent extension of conventional Voronoi models, in which cell boundaries are composed of straight Voronoi edges and circular arcs of fixed radius $\ell$.
When the line tension on cell-medium interfaces exceeds the tension on cell-cell contacts, we find that the model exhibits a strong time-step dependence in the fracture timescale of initially intact active clusters: decreasing $\Delta t$ can unphysically suppress cluster rupture events.
We trace this behavior to a divergence of detachment forces in the finite Voronoi model and introduce a simple regularization. {We then compare the finite Voronoi model's near-detachment mechanics to a deformable polygon model and propose two potential calibration strategies.
Finally, we examine the fracture--no-fracture transition in nonconfluent tissues and show that it is governed by detachment mechanics: the calibration can even determine the sign of the transition's dependence on cell shape.}
Our results show that, for studies focused on fracture or intercellular adhesion in nonconfluent monolayers, a physically motivated calibration of near-detachment mechanics in the finite Voronoi model is essential.
\end{abstract}

\maketitle
\end{CJK*}

\section{Introduction}

Both confluent tissues, in which cells densely pack and tile space, and nonconfluent cell populations, in which cells or clusters of cells are separated by gaps, are common in biology. Many systems transition between these two states.
A prominent example is epithelial-mesenchymal transition (EMT), in which epithelial tissues lose cohesion and give rise to dispersed, migratory cells or clusters~\cite{youssef2024epithelial, tripathi2020physics}.
Similarly, during cancer dissemination, tumor cell collectives can transition from a compact, confluent mass to a nonconfluent, invasive state~\cite{harper2016mechanism, cheung2016collective}.
Understanding such transitions is central to describing collective cell behavior, and these approaches require simulation methods that can correctly resolve the mechanics of both confluent and nonconfluent tissues. One popular and powerful approach for simulating confluent tissues is the active Voronoi model~\cite{bi2016motility}, in which cells have a polygonal shape given by the Voronoi tessellation of cell centers, and cells evolve under a mechanical energy that depends on this geometry. These models have been particularly successful in explaining the jamming transition of confluent monolayers, where cells tile space without gaps and neighbor exchanges occur through topological rearrangements~\cite{bi2015density, bi2016motility, chen2022activation}---but the commonly simulated active Voronoi model~\cite{bi2016motility} does not allow for simulating nonconfluent tissues. However, recent work has developed so-called ``finite Voronoi'' models, originating from the concept of free Dirichlet domains~\cite{graner1993can}, to extend the conventional Voronoi model to describe nonconfluent monolayers~\cite{bock2010generalized, schaller2005multicellular, teomy2018confluent, huang2023bridging}. Finite Voronoi model retain the advantages of Voronoi-based descriptions in characterizing cell shape, while remaining significantly less computationally demanding than approaches like phase-field models \cite{nonomura2012study, palmieri2015multiple, chiang2024multiphase, wang2025confinement}, subcellular element models~\cite{sandersius2008modeling, sweet2011modelling, sandersius2011emergent}, %
{or deformable polygon models and related variants~\cite{boromand2018jamming, lv2024active, weng2022convergent,kim2021embryonic}}, making it a promising candidate for simulating cohesive clusters, active fragmentation, and tissue rupture~\cite{huang2023bridging, teomy2018confluent}.

Here, we study the finite Voronoi model of Refs.~\cite{teomy2018confluent,huang2023bridging}, and demonstrate that in some circumstances the force required to detach two cells can diverge unphysically due to the imposed geometry of cell-cell contacts, preventing cells from rearranging. This is akin to---though more dramatic than---earlier results showing that the assumption of a strict Voronoi shape can eliminate the unjamming transition observed in vertex models~\cite{sussman2018no} and change the dynamics of heterotypic interfaces~\cite{lawson2024differences,yue2024scale}.  {Our results are part of the broader recognition of cusp-like, non-smooth forces at interfaces arising from topological interactions in cell mechanics~\cite{sussman2018soft}.}

We introduce the active finite Voronoi model in Sec.~\ref{sec:model}. We then demonstrate that when simulating tissue fracture, the results systematically and unavoidably depend on the time step used in numerical simulation, with fracture vanishing at small time steps (Sec.~\ref{sec:time-step}).
We then show that this pathology is not a numerical accident but a property of the finite Voronoi geometry. In Sec.~\ref{sec:doublet}, we analyze the simplest case of cell doublet detachment and show that the detachment force will diverge if there is an interfacial tension that differs between cell-cell and cell-medium interfaces.
Based on this analysis, we propose a regularization to address this divergence in Sec.~\ref{sec:regularization}.
In Sec.~\ref{sec:calibration}, we compare the finite Voronoi model with a deformable polygon model {that has the same energy function, but does not assume the strict Voronoi shape} and present two different calibration strategies in Secs.~\ref{sec:cal1} and \ref{sec:cal2}.
{Finally, in Sec.~\ref{sec:fracture_transition}, we study the fracture--no-fracture transition in nonconfluent tissues using the finite Voronoi model and show that choices of regularization and calibration of the detachment forces can change these phase diagrams quantitatively and qualitatively.}

\section{Model}\label{sec:model}
\begin{figure}
\includegraphics[width=0.48\textwidth]{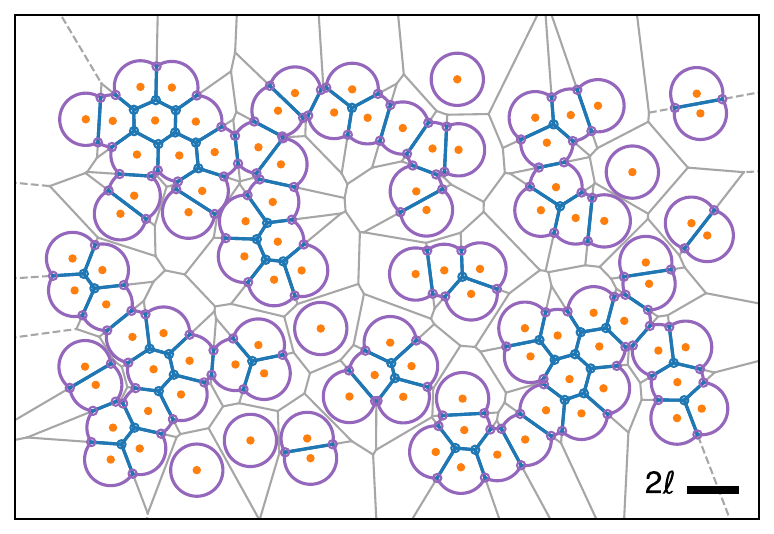}
\caption{\textbf{Illustration of the finite Voronoi (FV) model.} Cell center positions $\{\mathbf{r}_i\}$ are the orange points, and gray lines are the corresponding Voronoi diagram generated by $\{\mathbf{r}_i\}$ while the dashed lines represent rays of Voronoi edges that extend to infinity. 
Empty circles represent vertices $\{\mathbf{h}_m\}$ in the FV model: Blue circles are the triple junction vertices connecting three cells, while purple circles are the outer vertices $\mathbf{h}^{\textrm{out}}$ connecting two cells. 
Scale bar: maximum diameter $2\ell$.
}
\label{fig:finite_Voronoi}
\end{figure}

We consider $N$ cells in two dimensions, labeled by $i=1,\dots,N$, with cell centers at positions $\{\mathbf r_i\}$. 
In a conventional Voronoi model for confluent monolayers~\cite{bi2016motility}, Voronoi tessellations tile all of space, and the Voronoi region of cell $i$ is the set of points closer to $\mathbf{r}_i$ than to any other center.
In the finite Voronoi (FV) model, we assume that the cell cannot extend further than a distance $\ell$ away from its center. In this way, the shape of a cell is obtained by truncating the Voronoi region at a maximum radius $\ell$ from the cell center~\cite{teomy2018confluent,huang2023bridging}.
More formally, the domain of cell $i$ is the intersection of its Voronoi region with the disk $\{\mathbf x:|\mathbf x-\mathbf r_i|\leqslant \ell\}$.
As a result, FV boundaries are composed of (i) straight Voronoi segments where two cells are in contact and (ii) circular arcs of radius $\ell$ where the cell is exposed to the surrounding medium (Fig.~\ref{fig:finite_Voronoi}).
This construction allows two cells to separate once their centers move sufficiently far apart, creating gaps between cells without introducing additional degrees of freedom.%

The energy of the cells is defined similarly to that in the conventional Voronoi/vertex-based model: a quadratic elastic energy penalizing area deviation away from a preferred area, a quadratic energy penalizing cell perimeter, and line tensions that are potentially different between cell-cell and cell-medium interfaces~\cite{huang2023bridging,teomy2018confluent,chen2022activation}:
\begin{equation}
    E=\sum_i K_A(A_i-A_0)^2+K_PP_i^2 + \lambda^{(c)} P_i^{(c)}+\lambda^{(n)}P_i^{(n)},
\end{equation}
where $K_A$ and $K_P$ are elastic moduli for cell area $A_i$ and perimeter $P_i$, respectively, and $A_0$ is the preferred cell area. $\lambda^{(c)}$ and $\lambda^{(n)}$ are the cortical tensions for contacting edges (cell-cell) and non-contacting edges (cell-medium interface), and the total circumference $P_i=P_i^{(c)}+P_i^{(n)}$ is the sum of contacting length and non-contacting length.
Up to an additive constant, this can be rewritten in a ``preferred perimeter'' form:
\begin{equation}
    E=\sum_i K_A(A_i-A_0)^2 + K_P(P_i-P_0)^2 + \Lambda P_i^{(n)},
\end{equation}
where $P_0=-\lambda^{(c)}/2 K_P$ is the preferred perimeter, and $\Lambda \equiv \lambda^{(n)}-\lambda^{(c)}$ measures the tension difference between contacting and non-contacting edges, {reflecting the combined effects of cortical tension and cell-cell adhesion~\cite{manning2010coaction, maitre2012adhesion}.}

We assume that the cell centers are overdamped, so cell $i$ has one term in its velocity proportional to the force $-\nabla_i E$ on cell $i$ and one arising from self-propulsion~\cite{huang2023bridging},
\begin{equation}
    \dot{\mathbf{r}}_i = -\mu\bm{\nabla}_i E + v_0\mathbf{n}_i,
\end{equation}
where $\mu$ is the cell mobility, $v_0$ sets the self-propulsion speed, and $\mathbf{n}_i=(\cos\theta_i,\sin\theta_i)$ is the polarity direction---the direction the cell would travel in the absence of cell-cell interactions. The polarity undergoes rotational diffusion,
\begin{equation}
    \dot{\theta}_i=\sqrt{2D_r}\eta_i(t),
\end{equation}
with zero-mean, unit-variance
Gaussian white noise satisfying $\langle\eta_i(t)\rangle=0$ and $\langle\eta_i(t)\eta_j(t')\rangle=\delta_{ij}\delta(t-t')$.

We perform simulations of the active finite Voronoi (AFV) model using our own purpose-built \texttt{Python} package \texttt{PyAFV}~\cite{wang2026code}.
All equations are nondimensionalized using $K_A$
and a characteristic length scale $L$, e.g., $K_P$ is scaled by $K_A L^2$ and $\Lambda$ is scaled by $K_A L^3$~\footnote{{This corrects a typo in Eq.~(E4) of Ref.~\cite{teomy2018confluent}, where $\Lambda$ should be scaled by $K_AL^3$ rather than $K_AL^4$.}}.
Following Ref.~\cite{huang2023bridging}, we choose the length scale $L=\sqrt{A_0/\pi}$ so that the nondimensional preferred area is $A_0=\pi$.
Time is scaled by the mechanical relaxation time $1/\mu K_A L^2$. Unless otherwise stated, we set the scaled maximum radius $\ell=1$ and the scaled $K_P=1$ (see Table~\ref{tab:param}), and we vary the control parameters $P_0$ and $\Lambda$ to explore the detachment forces. Though we have implemented the model independently, our model assumptions are identical to those of Ref.~\cite{huang2023bridging}, except that we simulate with open boundaries rather than periodic boundaries.

\section{Divergence of the detachment forces}

\subsection{Tissue fracture timescale depends systematically on time step}\label{sec:time-step}

\begin{figure}
\includegraphics[width=0.49\textwidth]{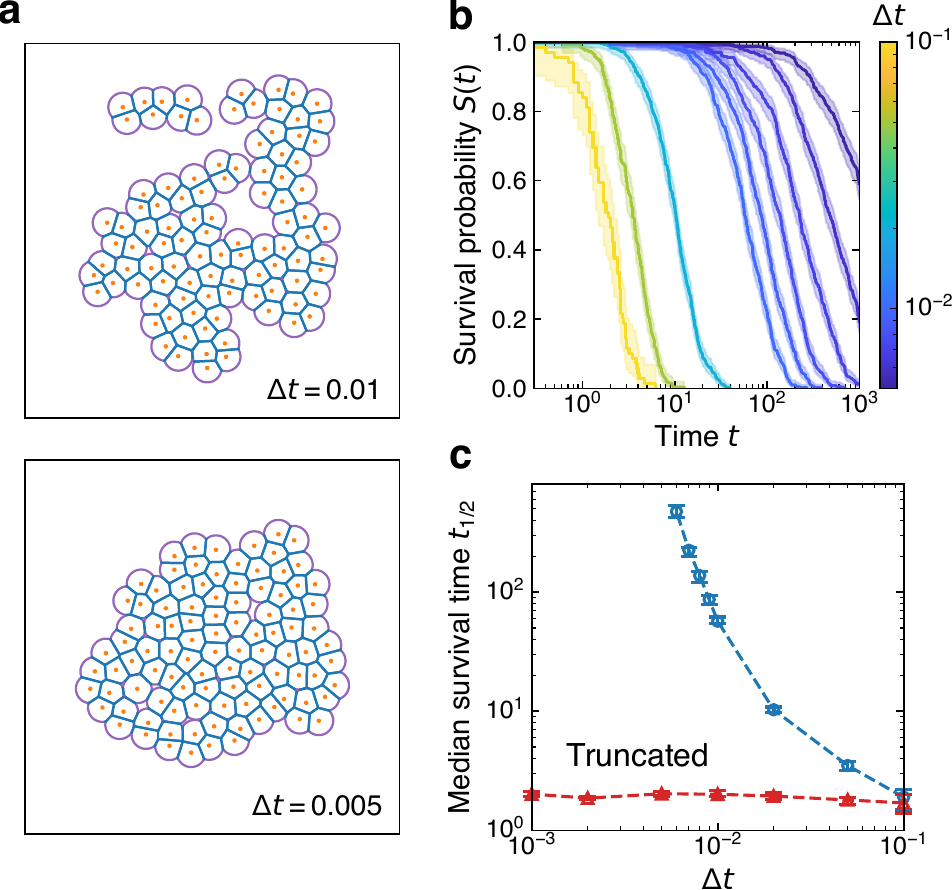}
\caption{\textbf{Tissue cohesion is strongly dependent on simulation timestep. a}, Simulation snapshots of an $N=100$ cell cluster at $t=100$ for time steps $\Delta t = 0.01$ (upper) and $\Delta t = 0.005$ (lower).
Each simulation was first evolved for $20$ time units with zero motility to reach a steady state, followed by $100$ time units of active dynamics.
\textbf{b}, Survival probability $S(t)$ of an initially intact monolayer with time steps $\Delta t=0.1$, $0.05$, $0.02$, $0.01$, $0.009$, $0.008$, $0.007$, $0.006$, and $0.005$.
\textbf{c}, Median survival time $t_{1/2}=S^{-1}(1/2)$ as a function of $\Delta t$; the cutoff is set to $\delta=0.45$ for the truncated curve.
For each value of $\Delta t$, we apply standard Kaplan-Meier survival analysis~\cite{kaplan1958nonparametric} to the results of $480$ independent simulations; error bars represent $95\%$ confidence intervals.
Parameters: number of cells $N=100$, initial packing fraction $\phi=0.5$, $\ell=1$, $K_P=1$, $A_0=\pi$, $P_0=4.8$, $\Lambda=0.1$, angular diffusion coefficient $D_r=1.33$, and active velocity $v_0=1.5$.
}
\label{fig:timestep}
\end{figure}

When performing a simple time-step convergence check of the AFV model, we find a strange phenomenon: When the tension difference %
{$\Lambda > 0$} %
and a smaller time step $\Delta t$ is used to evolve the dynamics, cells appear to be more adherent to each other, and fewer gaps form between cells.
As shown in Fig.~\ref{fig:timestep}a, we evolve a cluster of $N=100$ cells using different time steps, $\Delta t=0.01$ and $\Delta t=0.005$.
After a simulation of $100$ time units, the cluster evolved with the smaller $\Delta t$ is more compact than the one evolved with the larger $\Delta t$, and unlike the large-$\Delta t$ cluster, it has not fractured or formed holes.
{We have also confirmed that this behavior occurs in the code released by Ref.~\cite{huang2023bridging}.} One way to quantify cluster cohesion is to consider the time scale of rupture of the cluster---i.e., for an initially $N$-cell cluster, due to the random motility of individual cells, when does it rupture into two disconnected clusters?
{Such rupture events are commonly observed in experiments~\cite{prakash2021motility, duque2024rupture, wang2025confinement} and} can be measured using the survival probability $S(t)=\mathbb{P}(T>t)$, where we monitor the time $T$ of the \emph{first} rupture event~\cite{wang2026controlling, wang2025confinement}.
Figure~\ref{fig:timestep}b shows $S(t)$ for a range of $\Delta t$: As $\Delta t$ decreases, rupture is delayed, and the survival curves shift to longer times.
A convenient summary statistic is the median survival time $t_{1/2}=S^{-1}(1/2)$---the time at which the survival probability drops to one half---which grows rapidly as $\Delta t$ is reduced (blue curve in Fig.~\ref{fig:timestep}c).

Fig.~\ref{fig:timestep}b,c shows our qualitative visual analysis is correct: the cells become more adherent as the time step $\Delta t$ decreases. Why does this happen? Does this reflect an issue with numerical algorithms, such as an incorrect choice of time integration schemes, or is it something more fundamental? In the next section, we show that this is an unavoidable consequence of the underlying geometric assumptions of the model. %

\subsection{Forces between a cell doublet}\label{sec:doublet}

\begin{figure}
\includegraphics[width=0.48\textwidth]{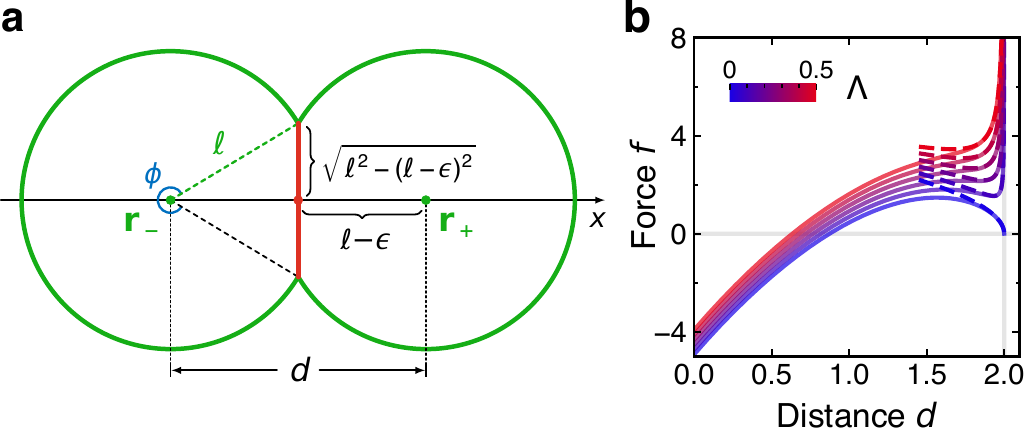}
\caption{\textbf{Cell doublet. a}, Schematic of a cell doublet with centers positioned symmetrically at $\mathbf{r}_\pm=\pm(\ell - \epsilon)\hat{x}$. The arc radius of each cell is $\ell$, and the distance between the two cell centers is given by $d = 2(\ell - \epsilon)$, thus the contact length (red) is $2\sqrt{\ell^2-(\ell-\epsilon)^2}$.
\textbf{b}, Force between a cell doublet as a function of their separation distance for $\Lambda=0,$ $0.1$, $0.2$, $0.3$, $0.4$, and $0.5$ (blue to red). Positive values $f>0$ correspond to attractive forces. Solid lines show Eq.~\eqref{eq:force}, and dashed lines are the asymptotic form for $d\to2\ell$ from Eq.~\eqref{eq:force_0}. Default parameters: $\ell=1$, $K_P=1$, $A_0=\pi$, and $P_0=4.8$.
}
\label{fig:doublet}
\end{figure}

To identify the origin of the divergence, we begin with the simplest detachment event: two identical cells that share a single contact.
As shown in Fig.~\ref{fig:doublet}a, we place the two cell centers symmetrically at $\mathbf{r}_\pm=\pm(\ell-\epsilon)\hat{x}$, so that the center-to-center distance is $d=2(\ell-\epsilon)$~\cite{teomy2018confluent}.
The parameter $\epsilon$ therefore measures how close the pair is to detachment: $\epsilon>0$ corresponds to a finite contact ($d<2\ell$), and $\epsilon\to0^+$ is the limit in which cells detach.
In this geometry, each cell boundary consists of a circular arc of length $P^{(n)}=\ell\phi$ together with a straight contact segment of length $P^{(c)}=2\sqrt{\ell^2-(\ell-\epsilon)^2}$, where $\phi=2\pi-2\operatorname{atan2}\bigl(\sqrt{\ell^2-(\ell-\epsilon)^2},\ell-\epsilon\bigr)$ is the angle spanning the region not in contact.

Because the doublet is symmetric, it is convenient to use the energy of a single cell:
\begin{equation}\label{eq:single_cell}
    E_\mathrm{s}=K_A(A-A_0)^2+K_P(P-P_0)^2+\Lambda P^{(n)},
\end{equation}
where the area and perimeter of each cell are
\begin{eqnarray}
    A&=&(\ell-\epsilon)\sqrt{\ell^2-(\ell-\epsilon)^2}+\frac{\ell^2\phi}{2},\nonumber\\
    P&=&2\sqrt{\ell^2-(\ell-\epsilon)^2}+\ell\phi.\nonumber
\end{eqnarray}
The (passive) interaction force between the two cells follows from the derivative of $E_\mathrm{s}$ with respect to the separation parameter,
$f=-\partial E_\mathrm{s}/\partial\epsilon$.
Evaluating this derivative gives (see Appendix~\ref{app:analyt} for details and some asymptotic limits)
\begin{eqnarray}
    f&=&4\sqrt{(2\ell-\epsilon)\epsilon}\left[K_A(A-A_0)+ K_P\dfrac{(P-P_0)}{2\ell-\epsilon}\right] \nonumber\\
    &&+ {\Lambda}\dfrac{2\ell}{\sqrt{(2\ell-\epsilon)\epsilon}},~~~~~~(\ell\geqslant\epsilon>0)\label{eq:force}
\end{eqnarray}
where in our sign convention $f>0$ corresponds to attractive forces (i.e., forces that tend to increase $\epsilon$). We see that the last term in Eq.~\eqref{eq:force} diverges as the two cells separate, $\epsilon \to 0^+$. This term comes from the diverging derivative of the non-contacting perimeter $P^{(n)}$ with $\epsilon$---in the finite Voronoi representation, small changes of the cell centers can lead to very large changes of perimeters.

The divergence of the detachment forces explains the time-step dependence of the rupture time scale: {when $\Lambda > 0$} cells \emph{should not be able to detach from one another} in the FV model of Refs.~\cite{huang2023bridging, teomy2018confluent}. The presence of detachment occurs only because {numerical simulation of the finite Voronoi model uses a} finite time step $\Delta t$, where the numerical evolution can allow the cell to ``skip'' the divergence to reach full cell-cell separation, where the force is zero. 
{It is possible for numerical errors to skip the divergence in part because the rapid increase in force is only apparent at distances very close to full separation (Fig.~\ref{fig:doublet}b)---if the cells move from distance $d\approx 1.95$ to over $2$ in a single time step, the divergence would not be felt. Given the value $v_0 = 1.5$ and a time step $\Delta t = 0.05$, this could be quite common.}
By decreasing the time step $\Delta t$, we are approaching the ``correct'' no-detachment limit where $t_{1/2}\to\infty$ (Fig.~\ref{fig:timestep}c). Clearly, this behavior is not desired in a model of cells that can detach from one another, and we attempt to regularize these divergences in the next section.

\subsection{Regularization by introducing a cutoff}\label{sec:regularization}

To regularize the divergent detachment force, it is useful to identify precisely where the divergence enters the force calculation in the active finite Voronoi model.
Given the cell-center positions of $N$ cells $\{\mathbf{r}_i\}$, the finite Voronoi structure produces a set of vertices $\{\mathbf{h}_m\}$ (including inner triple-junction vertices and outer vertices where straight edges meet circular arcs; see Fig.~\ref{fig:finite_Voronoi}).
Forces on cell centers follow from the chain rule:
\begin{equation}\label{eq:fx}
    {f}_{i,x}=-\frac{\partial E}{\partial x_i}=-\sum_m\frac{\partial E}{\partial \mathbf{h}_m}\cdot\frac{\partial \mathbf{h}_m}{\partial x_i},
\end{equation}
{where $x_i$ is the $x$ component of $\mathbf{r}_i$,} with an analogous expression for the $y$ component.

The divergence does \emph{not} originate from the inner vertices (circumcenters of Delaunay triangles), whose derivatives remain finite for generic configurations.
Instead, it comes from \emph{outer} vertices $\mathbf{h}^{\mathrm{out}}$ that connect a straight Voronoi segment to a circular arc.
For two cells $i$ and $j$ sharing an outer vertex, {there are two points} $\mathbf{h}_\pm^\mathrm{out}$ that lie at a distance $\ell$ from both centers, which are~\cite{teomy2018confluent}
\begin{equation}\label{eq:h_out}
    \mathbf{h}^{\textrm{out}}_\pm=\frac{\mathbf{r}_i+\mathbf{r}_j}{2}\pm\frac{\sqrt{4\ell^2-|\mathbf{r}_i-\mathbf{r}_j|^2}}{2|\mathbf{r}_i-\mathbf{r}_j|}(\mathbf{r}_i-\mathbf{r}_j)\times\hat{z},
\end{equation}
The $\pm$ reflects the two symmetric intersection points of the circles of radius $\ell$ centered at $\mathbf{r}_i$ and $\mathbf{r}_j$, lying on opposite sides of the line $ij$.
Their derivatives with respect to $x_i$ are given by
\begin{eqnarray}
    \frac{\partial \mathbf{h}^\textrm{out}_\pm}{\partial x_i}&=&\dfrac{\hat{x}}{2}\mp\frac{2\ell^2(x_i-x_j)}{\sqrt{4\ell^2-|\mathbf{r}_i-\mathbf{r}_j|^2}}\frac{(\mathbf{r}_i-\mathbf{r}_j)}{|\mathbf{r}_i-\mathbf{r}_j|^3}\times\hat{z}\nonumber\\
    &&\phantom{\frac{1}{2}}\mp\frac{\sqrt{4\ell^2-|\mathbf{r}_i-\mathbf{r}_j|^2}}{2|\mathbf{r}_i-\mathbf{r}_j|}\hat{y}.\label{eq:outer_derivative}
\end{eqnarray}
We can see that $\partial \mathbf{h}_\pm^{\mathrm{out}}/\partial x_i$ contains a term with $\sqrt{4\ell^2-|\mathbf{r}_i-\mathbf{r}_j|^2}$ in the denominator, which diverges as $|\mathbf{r}_i-\mathbf{r}_j|\to2\ell$.
This geometric divergence is the origin of the divergent detachment force in Eq.~\eqref{eq:force}.

Physically, $\sqrt{4\ell^2-|\mathbf{r}_i-\mathbf{r}_j|^2}=2\sqrt{\ell^2-(|\mathbf{r}_i-\mathbf{r}_j|/2)^2}$ is exactly the distance between the two outer vertices $\mathbf{h}^\textrm{out}_{\pm}$ generated by the pair $\{\mathbf{r}_i,\mathbf{r}_j\}$, as defined in Eq.~\eqref{eq:h_out}.
For the cell doublet shown in Fig.~\ref{fig:doublet}a, this quantity is simply the contact length $P^{(c)}$ (red).
Thus, a minimal practical regularization is to bound the vanishing denominator in Eq.~\eqref{eq:outer_derivative} by a small cutoff $\delta$:
\begin{equation}\label{eq:cutoff}
    \sqrt{4\ell^2-|\mathbf{r}_i-\mathbf{r}_j|^2} \to \max\left(\sqrt{4\ell^2-|\mathbf{r}_i-\mathbf{r}_j|^2},\delta\right).
\end{equation}
This prescription renders forces finite at detachment while leaving them unchanged for {$|\mathbf{r}_i-\mathbf{r}_j|<\sqrt{4\ell^2-\delta^2}=2\ell-(\delta^2/4\ell)+\mathcal{O}(\delta^4)$}, at the cost of introducing a new microscopic length scale $\delta$. This cutoff can be interpreted as {a threshold contact length below which the Voronoi shape assumption is faulty,} analogous to the finite edge-length criterion used to trigger T1 transitions in vertex models~\cite{chen2022activation, bi2015density, fletcher2014vertex, spahn2013vertex}. %

\begin{figure}
\includegraphics[width=0.48\textwidth]{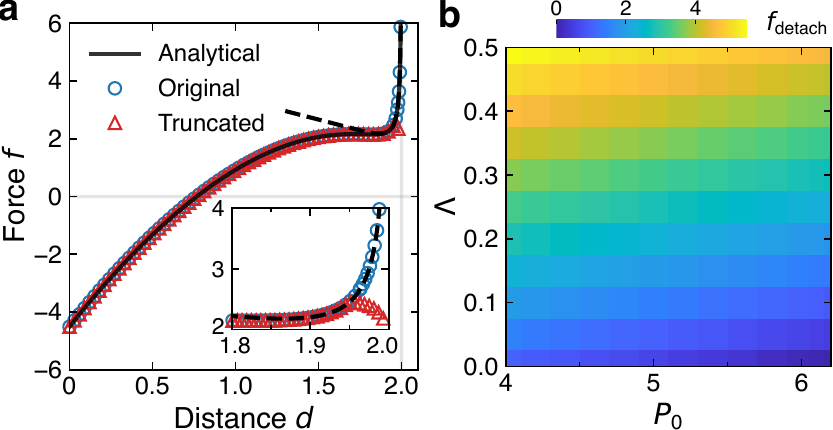}
\caption{\textbf{Regularization by introducing a cutoff. a}, Force between a cell doublet as a function of cell-center distance for $\Lambda=0.2$ and $P_0=4.8$. Empty circles and triangles represent forces obtained from simulations before and after implementing the truncation; analytical results are the corresponding curves in Fig.~\ref{fig:doublet}b. Inset: enlarged view around the cutoff $\delta=0.45$, which corresponds to $d_c=\sqrt{4\ell^2-\delta^2}\approx 1.95$.
\textbf{b}, Phase diagram for the final detachment force $f_\mathrm{detach}$ when $d=2\ell$ in the $P_0$--$\Lambda$ plane, where the cutoff $\delta=0.45$.
Default parameters: $\ell=1$, $K_P=1$, and $A_0=\pi$.}
\label{fig:truncated}
\end{figure}

From the shapes of the curves in Fig.~\ref{fig:doublet}b, for $\Lambda>0$ the force-distance relation exhibits a local minimum near $d\approx2\ell$, after which the force increases rapidly toward divergence.
For typical parameters, this rapid growth sets in around $d=1.95\ell$, corresponding to a threshold value $\delta=\sqrt{4\ell^2-d^2}\approx 0.45\ell$.
Applying the cutoff of Eq.~\eqref{eq:cutoff} with $\delta = 0.45$ (as $\ell = 1$), we successfully regularize the detachment forces around $d=2\ell$ (see Fig.~\ref{fig:truncated}a), and the truncation effect kicks in at $d_c\approx1.95$ as expected (inset of Fig.~\ref{fig:truncated}a). With this truncation, we eliminate the time-step dependence of the tissue fracture timescale (Fig.~\ref{fig:timestep}c, red curve).

In the finite Voronoi model, cell doublets are forced to detach when the cell-center distance $d$ exceeds $2\ell$. We extract the final detachment forces $f_\textrm{detach}$ at $d=2\ell$ numerically, and we find that, after introducing the regularization, this force is no longer divergent even when sweeping through the parameter space of $P_0$ and $\Lambda$ (Fig.~\ref{fig:truncated}b). {The detachment forces in Fig.~\ref{fig:truncated}b increase strongly with $\Lambda$, as we would expect---increasing $\Lambda$ corresponds to a higher energy cost for the cell-medium interface, increasing the cost for separation. We also see that increasing $P_0$ while holding other parameters constant weakly decreases the detachment force. This is consistent with, e.g., looking at the analytical detachment force in Eq.~\eqref{eq:force} at the distance at which regularization kicks in, $\epsilon_c = \ell - d_c/2$.}

\begin{figure}
\includegraphics[width=0.48\textwidth]{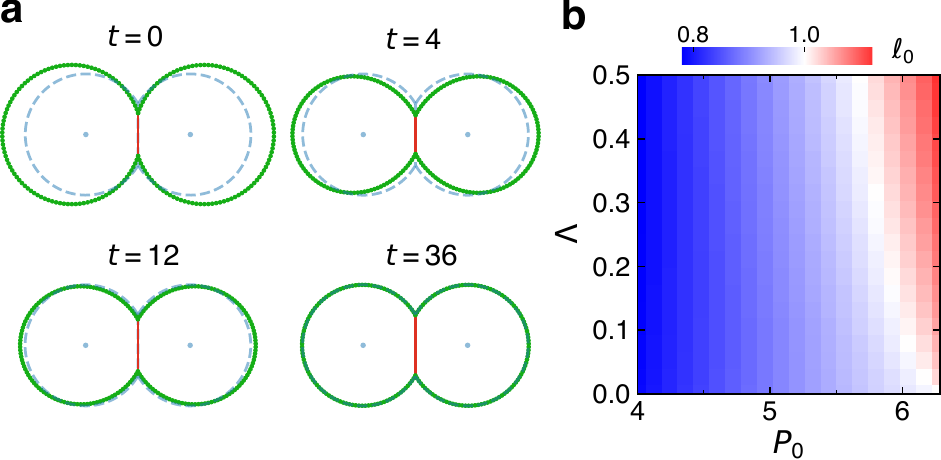}
\caption{\textbf{Steady states of deformable polygon model. a}, Simulation snapshots of how the cell doublet relaxes from its initial shape to the steady state in DP model. Blue shapes indicate the optimal $(\ell_0, \epsilon_0)\approx(0.87, 0.12)$ computed from Eq.~\eqref{eq:optimal} for $P_0=4.8$ and $\Lambda=0.2$.
\textbf{b}, Phase diagram for optimal $\ell_0$ in the $P_0$--$\Lambda$ plane.
Default parameters: $K_P=1$, $A_0=\pi$.
}
\label{fig:steady_state}
\end{figure}

\begin{figure*}
\includegraphics[width=0.75\textwidth]{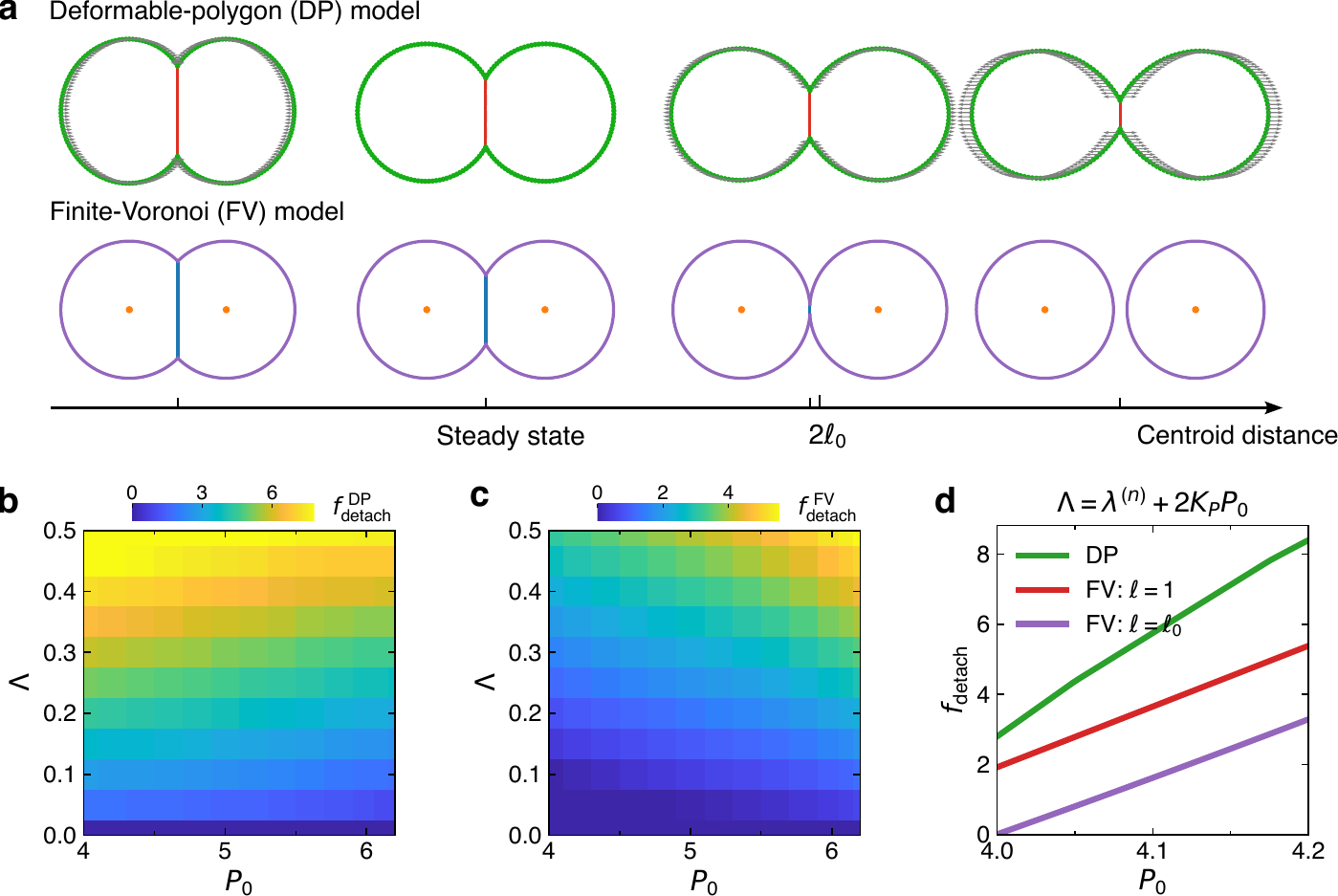}
\caption{\textbf{Detachment of cell doublets in the finite Voronoi model (FV) and deformable polygon (DP) models. a}, Snapshots of a cell doublet for both deformable polygon model and finite Voronoi model at different centroid distances.
Preferred perimeter is set to $P_0=4.8$ and $\Lambda=0.2$, yielding the optimal $(\ell_0, \epsilon_0)\approx (0.87, 0.12)$ which has been used to set the maximum radius $\ell=\ell_0$ in FV model. The four centroid distances from left to right are $1.374$ (compressed), $1.550$ (steady state), 1.735 (approaching detachment threshold $2\ell_0$ in FV), and $1.912$ ($>2\ell_0$).
\textbf{b}, The detachment forces $f_\textrm{detach}^\textrm{DP}$ for the DP model where contact length $P^{(c)}$ of the cell doublet approaches zero.
\textbf{c}, The detachment forces $f_\textrm{detach}^\textrm{FV}$ for the FV model when $d=2\ell$, where we have set the maximum radius to the steady-state radius from the DP model, i.e., $\ell=\ell_0$, and the threshold value $\delta=0.45$.
\textbf{d}, Detachment forces along the path $\Lambda=\lambda^{(n)}+2K_P P_0$ in the previous phase diagrams, with the line tension on cell-medium interfaces fixed at $\lambda^{(n)}=-7.9$.
Default parameters: $K_P=1$, $A_0=\pi$.}
\label{fig:detachment}
\end{figure*}

\begin{figure}
\includegraphics[width=0.49\textwidth]{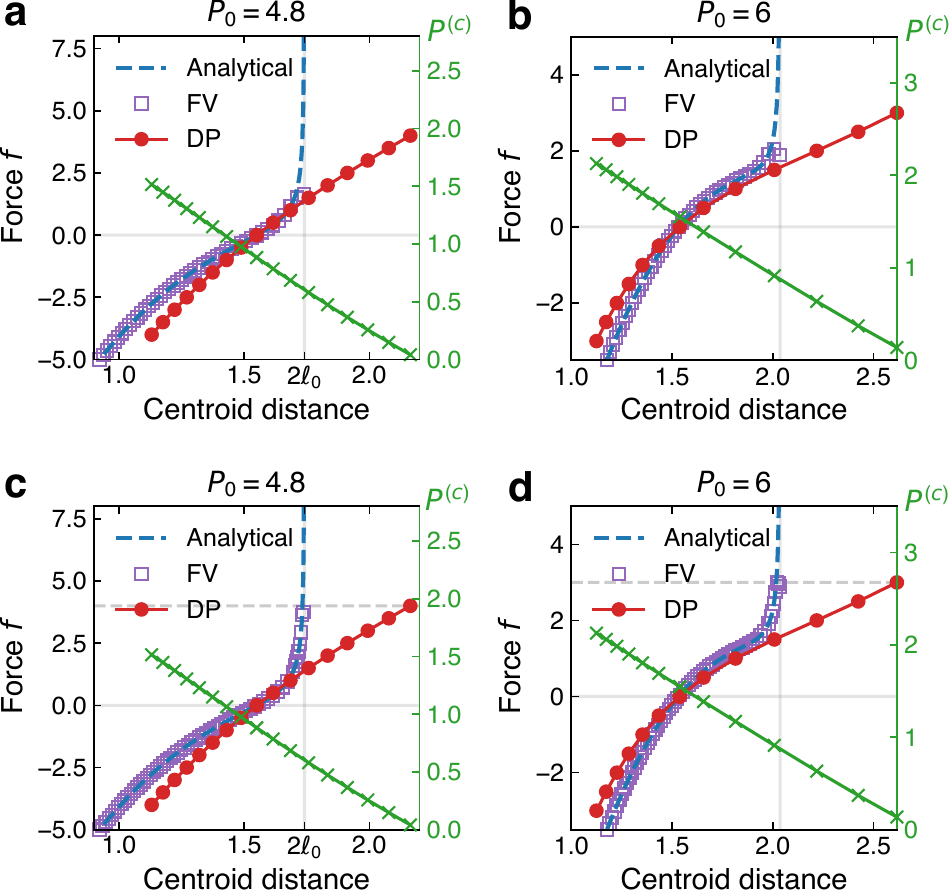}
\caption{\textbf{Calibrating the finite Voronoi model (FV) to a deformable polygon (DP) model.}
\textbf{a},\textbf{b} show how the forces in both models and also the contact length $P^{(c)}$ in DP model (green) vary with increasing centroid distance for two values of $P_0$ ($4.8$ and $6$), using calibration strategy 1 with $\ell=\ell_0$ and fixed $\delta=0.45$.
\textbf{c},\textbf{d}, Calibrate FV model by matching the detachment forces $f_\textrm{detach}$ to DP model with varying cutoff $\delta$.
Gray dashed lines indicate the match of detachment forces in both models.
Default parameters: $K_P=1$, $A_0=\pi$, and $\Lambda=0.2$.}
\label{fig:calibrate}
\end{figure}

\section{Calibration with a deformable polygon model}\label{sec:calibration}

The origin of the divergence we found in the finite Voronoi model is the assumption that, even as the cells are almost completely separated, the cells remain circular with a flat contact. %
During detachment, however, real cells exhibit appreciable shape deformation~\cite{smeets2019effect, vangheel2026rigidity, chu2005johnson, byers1995role, rozema2025remodeling,eckert2025cell}.
While we are able to regularize the divergence by introducing a new scale $\delta$, we want to better understand how to set this value. To calibrate {our regularized} finite Voronoi model, we compare it against a deformable polygon (DP) model that has the same energy function but in which cell boundaries are explicitly represented by polygons with large numbers of vertices that can deform continuously~\cite{boromand2018jamming, lv2024active, weng2022convergent}---i.e., the same energetics as our FV model but without the Voronoi geometry assumption. During calibration, we focus on passive mechanical forces and set $v_0=0$.

\subsection{Matching passive steady states between the deformable polygon model and finite Voronoi model}\label{sec:DP_model}

In the DP model, each cell is represented by a closed polygon with an ordered set of vertices $\{\mathbf h_{m}\}_{m=1}^M$ (Appendix~\ref{app:DP}).
We use the same energy as in the FV model: quadratic area and perimeter elasticity together with distinct line tensions on cell-cell contacts and cell-medium boundary.
However, the degrees of freedom are not the cell centers, but the vertices on the cell boundaries.
Vertex dynamics are evolved by overdamped relaxation:
\begin{equation}
    \frac{\mathrm{d}\mathbf{h}_m}{\mathrm{d}t}=-\mu\frac{\partial E}{\partial\mathbf{h}_m}. \label{eq:dp_evolve}
\end{equation}

To place the two models on a comparable footing, we first choose parameters so that the steady states of passive cell doublets match between the models. Minimizing the energy in the DP model using Eq.~\eqref{eq:dp_evolve}, we see that the minimal-energy state is, as the finite Voronoi model assumes, generally a pair of circular arcs with a straight boundary between them. However, the resulting steady-state radius of the arcs is the value that minimizes the energy, which will depend on the mechanical parameters of the model---and not the default $\ell=1$ used in the FV model. To match the FV and DP models, we must vary $\ell$ when we vary the FV model parameters, choosing the value that minimizes the energy. In this case, as long as the steady state of the DP model fits the finite Voronoi assumption, the FV model should recover the same steady state. Therefore, in our calibration process, we find the minimal-energy separation between cell centers and the minimal-energy value of $\ell$,
\begin{equation}\label{eq:optimal}
(\ell_0,\epsilon_0)=\operatorname*{arg\,min}_{\ell\geqslant\epsilon>0} E_\mathrm{s}(\ell,\epsilon),
\end{equation}
and then set the FV radius to $\ell=\ell_0$. We see that in the absence of external forcing, FV and DP predict the
same relaxed doublet shape (Fig.~\ref{fig:steady_state}a).
The values of the steady-state radius $\ell_0$ for different sets of $(P_0,\Lambda)$ are shown in Fig.~\ref{fig:steady_state}b. We see that $\ell_0$ increases with the preferred perimeter $P_0$, while $\Lambda$ has a relatively weaker effect on $\ell_0$.

\subsection{Force-distance curves in the deformable polygon model}

We want to find the relationship of force between cell pairs and distance in the DP model, akin to the values predicted for the finite Voronoi model in Fig.~\ref{fig:doublet}b. To do this, we apply equal and opposite forces to each cell in an initially relaxed doublet in the DP model, and measure the resulting displacement of the cell centroids $d$. (Note that to be consistent in our sense of cell-cell separation between the DP and FV model, here and throughout this section, we use $d$ to denote the centroid-centroid separation of both the DP and FV cells---cells in the DP model are not perfectly circular and do not have a ``center'' defined as in the FV model.) %
We distribute the force on one cell uniformly over all vertices of each cell to avoid spurious torques~\cite{smeets2019effect}, as shown in Fig.~\ref{fig:detachment}a.
We then record the centroid separation and contact length $P^{(c)}$ (red segments in Fig.~\ref{fig:doublet}a) as functions of the external force $f$. We define the detachment force $f_{\textrm{detach}}^\textrm{DP}$ as the force at which the contact length falls below the typical distance $\ell_c=2\pi\ell_0/M$ between neighboring vertices (see Appendix~\ref{app:DP}).
{This detachment force increases strongly with $\Lambda$ as expected, and decreases weakly with $P_0$ (Fig.~\ref{fig:detachment}b).}
If we choose parameters of the FV model such that the steady states of the FV and DP models match (i.e., $\ell$ is the energy-minimizing $\ell_0$), we find that the FV detachment force exhibits the opposite trend with $P_0$: for fixed $\Lambda$, increasing $P_0$ leads to larger $f_\textrm{detach}^\textrm{FV}$ (Fig.~\ref{fig:detachment}c), in contrast to the behavior in the DP model (Fig.~\ref{fig:detachment}b) and in the FV model with a fixed maximum radius ($\ell=1$, Fig.~\ref{fig:truncated}b).

{What behavior should we expect for detachment forces' dependence on $P_0$? In our rescaled units, $P_0=-\lambda^{(c)}/2K_P$, so increasing $P_0$ typically corresponds to stronger cell-cell adhesion (i.e., making cell-cell line tension $\lambda^{(c)}$ more negative). Stronger cell-cell adhesion must increase the detachment force, so at first glance, it initially appears that only Fig.~\ref{fig:detachment}c exhibits the physically correct trend.
However, $P_0$ and $\Lambda=\lambda^{(n)}-\lambda^{(c)}$ both depend on the more fundamental parameter $\lambda^{(c)}$.
If $\lambda^{(n)}$ is held fixed and only $\lambda^{(c)}$ is varied, increasing adhesion corresponds to moving along the line $\Lambda=\lambda^{(n)}+2K_P P_0$ in the $\Lambda$--$P_0$ plane.
Along this path, the detachment forces in both the finite Voronoi model and the deformable polygon model increase as cell-cell adhesion is made stronger (Fig.~\ref{fig:detachment}d).}

We compute the force-distance curve in the DP model and track the contact length as the cells are pulled apart from one another. We show these curves in Fig.~\ref{fig:calibrate}a,b for two different values of $P_0$. Here $f>0$ corresponds to a pull-off force dipole while $f<0$ corresponds to a pushing force dipole.
The contact length $P^{(c)}$ of the cell doublet {decreases roughly linearly with} the cell-centroid distance, and in the pulling limit ($f>0$) the force for the DP model increases with centroid distance almost linearly, consistent with Ref.~\cite{smeets2019effect}, where cells are modeled as three-dimensional elastic triangulated shells. %

{Comparing the analytical curves of the FV model and the DP simulation results in Fig.~\ref{fig:calibrate}a,b, we find two key results: 1) for small perturbations away from the relaxed state (zero force), the DP and FV models agree well, and 2) the DP model does not have a divergent detachment force, staying attached over a much larger cell-cell distance than the FV model. We also see by looking at the cell doublet shapes in Fig.~\ref{fig:detachment}a that at larger forces, the finite Voronoi shape assumption starts to fail---cells can have centroids separated by more than $2\ell_0$ without losing contact, and at higher forces cells are no longer well described by circular arcs (rightmost panel of Fig.~\ref{fig:detachment}a).} 

{We have found that, simply by choosing the energy-minimizing $\ell_0$, the small forces near the equilibrium are in agreement between FV and DP models. However, the detachment forces are clearly not in good agreement. Next, we present two potential approaches to address this.}

\subsection{Calibration strategy 1: setting a critical length} \label{sec:cal1}

As a first calibration strategy for near-detachment behavior, we choose the energy-minimizing value $\ell = \ell_0$ and adopt a fixed value of $\delta$, using the default value $\delta=0.45$ as in Fig.~\ref{fig:truncated}. {As discussed above, choosing a fixed value of $\delta$ corresponds to the idea that there is a minimal contact length beyond which cell-cell forces cannot increase, or equivalently a maximal cell-cell separation.}
{We choose the value $\delta = 0.45$ to capture a reasonable scale at which the force-distance curve flattens as in Fig.~\ref{fig:truncated}a---in Strategy 1 we do not use any information from the DP model to set this value.}
After setting $\ell=\ell_0$ and applying this fixed cutoff to truncate the rapidly growing force near detachment $d\approx2\ell_0$ in the FV model,
the simulation results (empty squares in Fig.~\ref{fig:calibrate}a,b) follow the analytical curves but lack the divergence near $2\ell_0$, as expected. Therefore, consistent with the comparison in the previous section between the analytical FV results and the DP model, we observe reasonable agreement between the FV and DP force-distance relations for $d\leqslant 2\ell_0$, with only a few deviations at the smallest cell-cell distances where cells are highly compressed. However, the agreement in the force-distance curves necessarily means that the detachment forces differ between DP and FV in Strategy 1---since cells remain connected even when $d>2\ell_0$ in the DP model, the DP model exhibits a much larger detachment force.

\begin{figure*}
\includegraphics[width=0.99\textwidth]{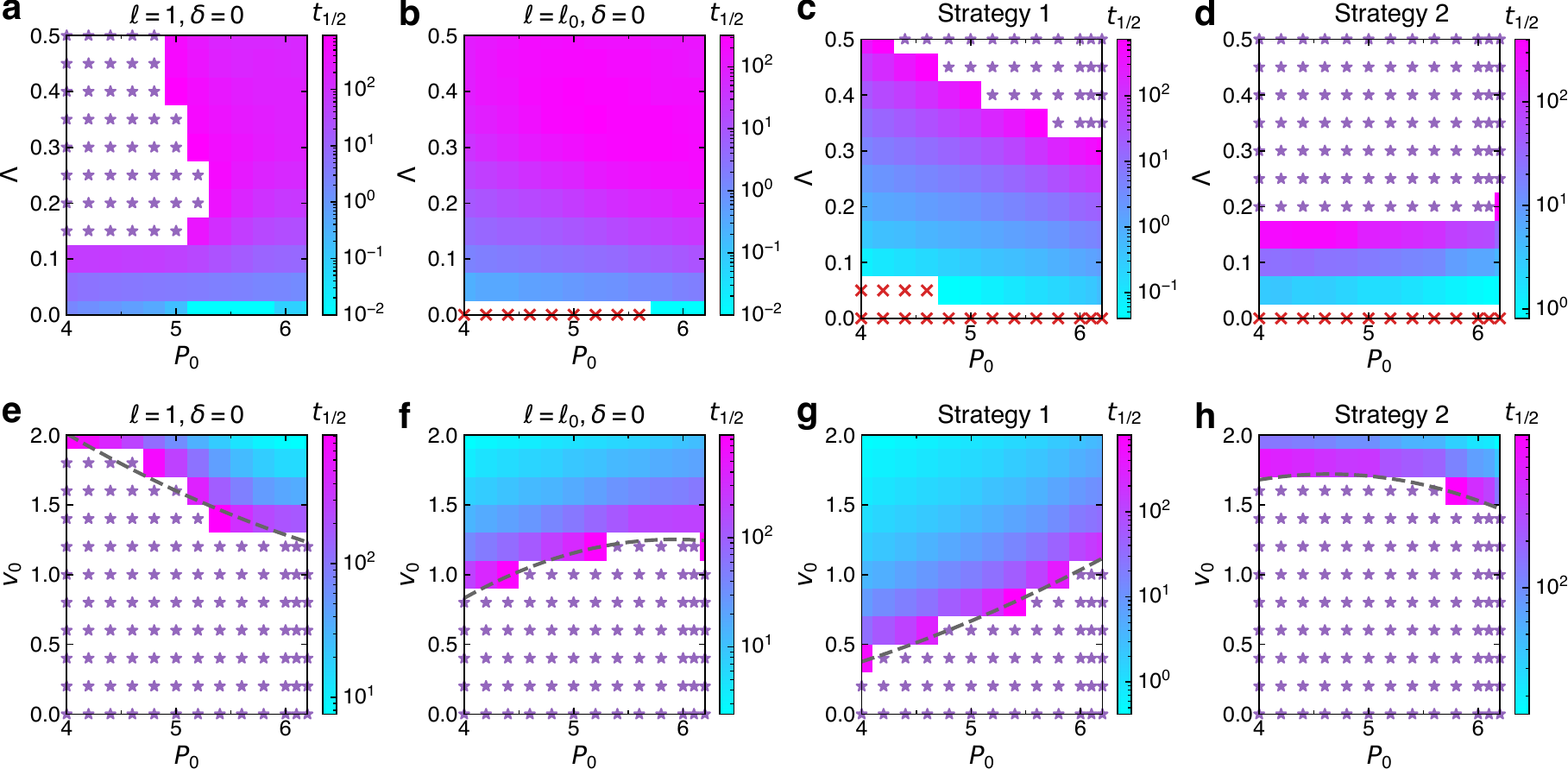}
\caption{{\textbf{Fracture--no-fracture transition characterized by the median survival time $t_{1/2}$.}
\textbf{a}--\textbf{d}, Phase diagrams of $t_{1/2}$ in the $P_0$--$\Lambda$ plane at fixed cell motility $v_0=1.5$.
\textbf{e}--\textbf{h}, Corresponding phase diagrams in the $P_0$--$v_0$ plane at fixed $\Lambda=0.2$.
\textbf{a},\textbf{e} show the results without regularization or calibration ($\ell=1, \delta=0$).
\textbf{b},\textbf{f} show the results after calibrating the radius $\ell$ to the deformable polygon model $\ell=\ell_0$, while leaving the divergence unregularized ($\delta=0$).
\textbf{c},\textbf{g} exhibit the results under calibration strategy 1, with $\ell=\ell_0$ and a fixed cutoff $\delta=0.45$.
\textbf{d},\textbf{h} show the results under calibration strategy 2, with $\ell=\ell_0$ and $\delta$ calibrated by matching the detachment forces of the FV and DP models.
Each data point of $t_{1/2}$ in the phase diagrams is obtained by computing the survival probability from the results of $480$ independent simulations; red crosses ($t_{1/2}=0$) indicate tissues are already ruptured at the start (after initial relaxation time); purple stars indicate that $t_{1/2}$ exceeds the total simulation time $T_\mathrm{tot}$; gray dashed lines in \textbf{e}--\textbf{h} are guides to the eye.
Default parameters: $K_P=1$, $A_0=\pi$, $D_r=1.33$.}}
\label{fig:phase_diagrams}
\end{figure*}

\subsection{Calibration strategy 2: matching detachment forces}\label{sec:cal2}

As discussed above, the previous strategy aligns the FV and DP models only for $d\leqslant 2\ell_0$, and this results in the FV model requiring a much smaller pull-off force to detach two cells.
Thus, we also consider an alternative calibration based on the detachment force $f_\textrm{detach}$.
In this approach, we set $\ell$ to be the energy-minimizing $\ell_0$ and adjust $\delta$ such that the detachment force in the FV model matches that of the DP model (Fig.~\ref{fig:detachment}b), i.e., $f_\textrm{detach}^\textrm{FV}=f_\textrm{detach}^\textrm{DP}$.
We numerically compute the value of $d_c$ required for Eq.~\eqref{eq:force} to equal $f_\textrm{detach}^\textrm{DP}$, and obtain the corresponding $\delta_c$. The results are shown in Fig.~\ref{fig:calibrate}c,d, where the horizontal dashed lines indicate the matched detachment forces. {A caveat of this calibration strategy is that it reintroduces the sharply increasing forces near detachment, thereby potentially requiring finer time stepping to maintain numerical stability in simulations.}

{
\section{Fracture--no-fracture transition in nonconfluent tissues}\label{sec:fracture_transition}
}

{
A key question for nonconfluent tissues is whether an initially cohesive cluster of cells stays together or active motility pulls it apart into dispersed pieces, as happens in collective cancer invasion~\cite{cheung2016collective, wang2025confinement} and \textit{Trichoplax} reproduction~\cite{prakash2021motility}.
This fracture--no-fracture transition can be characterized by the median survival time $t_{1/2}$, which quantifies the characteristic fracture timescale, as explored in Fig.~\ref{fig:timestep}b,c.
We expect this transition to be closely related to the clustered-to-dispersed transition studied in Ref.~\cite{huang2023bridging}, where the size of the largest connected cluster is used as the order parameter---a tissue that remains a single giant cluster has, by definition, not fractured.
In the finite Voronoi model, however, this transition is set by the near-detachment mechanics: a cluster can only disperse by breaking cell-cell contacts, and we found that the force needed to break a contact diverges in the absence of our regularization.
How does the fracture--no-fracture transition depend on the way this divergence is treated?
Figure~\ref{fig:phase_diagrams} shows $t_{1/2}$ across the $(P_0,\Lambda)$ and $(P_0,v_0)$ planes for four cases: (i) the original unregularized, uncalibrated model of Ref.~\cite{huang2023bridging} ($\ell=1$, $\delta=0$); (ii) using the energy-minimizing cell size but keeping the divergent detachment forces ($\ell=\ell_0$, $\delta=0$); (iii) calibration strategy 1 using a fixed cutoff ($\ell=\ell_0$, $\delta=0.45$); and (iv) calibration strategy 2 ($\ell = \ell_0$, $\delta$ chosen by matching the detachment forces to the deformable particle model).
}

{
The first two columns in Fig.~\ref{fig:phase_diagrams} are unregularized and not physical predictions.
With $\delta=0$ the detachment force diverges, so a cluster ruptures only because a finite time step lets its cells skip past the divergence, and the resulting transition line should shift with $\Delta t$.
We plot them at our default $\Delta t=0.01$ to show what the unregularized protocol produces---this is the same time step and approach used in  Ref.~\cite{huang2023bridging}, and our results should match theirs.}

{
In standard active Voronoi and vertex models, it is well established that a larger preferred perimeter $P_0$ leads to more fluid-like behavior \cite{bi2015density,bi2016motility}. 
For the unregularized, uncalibrated model of Ref.~\cite{huang2023bridging}, we find that, at fixed $\Lambda$, increasing $P_0$---corresponding to a more fluid-like state with a lower line tension $\lambda^{(c)}$ for contacting edges---facilitates fracture (Fig.~\ref{fig:phase_diagrams}a), consistent with the analytic result for the detachment force in Eq.~\eqref{eq:force}, which decreases with increasing $P_0$.
Regarding the line-tension difference, we would expect increasing $\Lambda$ to suppress fracture because it effectively increases $\lambda^{(n)}$, thereby penalizing the creation of non-contacting free boundaries. However, Fig.~\ref{fig:phase_diagrams}a shows that $t_{1/2}$ deceases slightly at very large $\Lambda$.
We attribute this nonmonotonic behavior to a numerical artifact of the unregularized model: we have observed that at large $\Lambda$, there can be such large forces between cells that cells can be ejected from the tissue monolayer unphysically within a single simulation time step.}
{In Fig.~\ref{fig:phase_diagrams}e, we observe that in the $(P_0, v_0)$ plane, with $\ell=1$, the motility needed to fracture the cluster \emph{decreases} with $P_0$, the same trend reported in Ref.~\cite{huang2023bridging}.}

{For the unregularized model of Fig.~\ref{fig:phase_diagrams}a,e, increasing $P_0$ generally increases the likelihood of fracture---consistent with the simulations of Ref.~\cite{huang2023bridging} with the unregularized model. This trend, however, depends critically on the assumptions made about $\ell$ and $\delta$. If we instead calibrate $\ell$ to the energy-minimizing $\ell_0$ (Fig.~\ref{fig:phase_diagrams}b,f), the dependence on cell shape $P_0$ \emph{reverses}: increasing $P_0$ now makes the cluster harder to break apart. (We note the numerical artifact for unphysical fracture at large $\Lambda$ remains present in Fig.~\ref{fig:phase_diagrams}b). This trend with $P_0$ is consistent with the increase of regularized detachment force with $P_0$ in the FV model with $\ell = \ell_0$ (Fig.~\ref{fig:detachment}c). 
Thus, even in the unregularized model ($\delta=0$) at fixed $\Delta t$, the choice of maximum radius alone can qualitatively reshape the fracture--no-fracture transition.
}

{
Regularizing the divergence removes the time-step artifact and turns the transition diagram into a well-defined physical prediction (the last two columns in Fig.~\ref{fig:phase_diagrams}).
Under Strategy~1, holding $\Lambda$ and $v_0$ constant, a larger preferred perimeter $P_0$ makes clusters harder to rupture (Fig.~\ref{fig:phase_diagrams}c,g).
This differs from Ref.~\cite{huang2023bridging}, where clusters with larger $P_0$ were more likely to be in dispersed states.
We attribute this discrepancy not to the choice of cutoff $\delta$, but to calibrating the maximum cell radius $\ell$ to its energy-minimizing value $\ell_0$, as in the change from Fig.~\ref{fig:phase_diagrams}a,e to Fig.~\ref{fig:phase_diagrams}b,f.
Moreover, regularization eliminates the unphysical fracture observed at large $\Lambda$: increasing $\Lambda$ now consistently makes clusters more resistant to fracture (Fig.~\ref{fig:phase_diagrams}c).}
{This strategy uses a fixed value of $\delta=0.45$. Changing the value of $\delta$ will quantitatively alter the detachment forces, which will influence the fracture timescales, but not affect the trends of fracture time with varying parameters (Fig.~\ref{fig:SM_vary_delta}).}%

{
Under Strategy~2, as noted in Sec.~\ref{sec:cal2}, the sharply increasing forces near detachment will require finer timestepping. To obtain reliable fracture statistics, we therefore gradually reduce the time step $\Delta t$ from the default value of $0.01$ to $0.002$ until the results converge (see Fig.~\ref{fig:SM_vary_dt}).
As shown in Fig.~\ref{fig:phase_diagrams}d,h, matching the detachment forces generally makes clusters harder to disperse than under Strategy~1 (more points appear as stars, i.e., the half-time to fracture exceeds simulation time) because the detachment forces become much larger after matching.
Moreover, the transition lines between no-fracture and fracture states differ from those in Fig.~\ref{fig:phase_diagrams}c,g: the fracture timescale depends only weakly on the preferred perimeter $P_0$ now, consistent with the detachment-force phase diagram in Fig.~\ref{fig:detachment}b, whereas Strategy~1 corresponds to Fig.~\ref{fig:detachment}c.
}

{
In general, the trend with respect to cell motility remains the same as that reported in Ref.~\cite{huang2023bridging}: increasing the motility $v_0$ promotes fracture.
Also, after introducing regularization $\delta$, increasing $\Lambda$ consistently suppresses fracture.
By contrast, the dependence on cell shape does not persist: how the near-detachment mechanics are calibrated sets the sign of the $P_0$ trend, and hence the shape of the fracture--no-fracture boundary.
Therefore, the fracture--no-fracture transition of a nonconfluent active tissue cannot be predicted from the finite Voronoi model without first establishing these mechanics.} %

\section{Discussion}\label{sec:discussion}

Our results show that the finite Voronoi model contains a pathological behavior when cell-cell and cell-medium tensions differ: The force required to separate two cells diverges, so fracture in active clusters becomes spuriously controlled by the numerical time step rather than by physical parameters.
We trace the origin of this divergence analytically by analyzing a two-cell system and introduce a cutoff threshold that removes the divergence and restores a well-defined fracture timescale. We then compare the finite Voronoi model with a deformable polygon model and propose two calibration strategies, providing a practical framework for regularizing and calibrating the model in studies of tissue fracture and cell separation. {We found that choices of regularization and calibration can qualitatively change the fracture--no fracture phase diagram.}

The cutoff regularization introduced in Eq.~\eqref{eq:cutoff} removes the time step dependence of the AFV model, but it modifies near-detachment mechanics by introducing a new length scale $\delta$.
We think of $\delta$ as a truncation threshold for the contact length $P^{(c)}$: once the contact shrinks below $\delta$, the intercellular force no longer increases.
However, the precise value of this threshold is both model- and parameter-dependent, and cannot be uniquely fixed within the FV framework.
Recent work using finite-element descriptions of cell mechanics has emphasized that tissue-level phenomena such as jamming can depend sensitively on the force required to separate neighboring cells~\cite{vangheel2026rigidity}, suggesting that the detachment force should be viewed as an important physical control parameter rather than a purely numerical detail.
From this perspective, the regularization parameter $\delta$ provides a practical way to tune the FV model so that its near-detachment mechanics match those of more detailed descriptions.
For the two calibration strategies introduced in Sec.~\ref{sec:calibration}, the appropriate choice depends on the physical regime of interest.
If detachment events are rare---for example, when the tissue remains largely in a solid-like state---Strategy~1 is typically sufficient, since it already reproduces the force-distance relationship well near the steady-state configuration.
By contrast, when frequent cell rearrangements or fracture events occur, Strategy~2 is more appropriate because it matches the detachment forces between the FV and DP models.
More generally, the cutoff $\delta$ can also be calibrated to match detachment forces obtained from more detailed models, such as those in Refs.~\cite{boromand2018jamming, lv2024active}, or even from experimental measurements of cell-cell pull-off forces.

The divergence of detachment forces in the finite Voronoi model may affect the results of Ref.~\cite{huang2023bridging} {as their primary simulations all had $\Lambda > 0$, where we observe divergences}.
First, because no force cutoff was implemented in Ref.~\cite{huang2023bridging}, clusters could fracture only due to the finite simulation time step $\Delta t$. Although the numerical dynamics with finite $\Delta t$ will act as an effective regulator similar to the threshold $\delta$ introduced here, we think the use of the regularizing $\delta$ avoids possible complications in which varying numerical details like $\Delta t$ can have spurious physical effects. We also {observe relevant quantitative differences between implicitly regularizing with a finite $\Delta t$ and our approach (Fig.~\ref{fig:phase_diagrams})}. %
Secondly, choices about the calibration of $\ell$ and regularization can qualitatively change results. Using a fixed $\ell$ rather than the steady-state radius $\ell_0$ from the DP model reverses the trend of fracture behavior when varying $P_0$ (Figs.~\ref{fig:truncated}b and \ref{fig:detachment}c).
{Interestingly, in our model using a varying $\ell_0$ and calibrating to the detachment forces (Fig.~\ref{fig:phase_diagrams}h), which we think is the most reasonable approach, we do see the same qualitative trends in the effect of $P_0$ that are observed by Ref.~\cite{huang2023bridging} in their phase diagrams (e.g., their Fig.~2b). However, we believe the origin of these behaviors may be subtler than expected, and they depend on our calibration choices.}
{Therefore, different calibration strategies may alter the clustered-dispersed transition diagram reported in Ref.~\cite{huang2023bridging} and may also shift the jamming-transition boundary, whereas the confluency transition (whether small interstitial gaps exist between cells~\cite{rustarazo2026adhesion}), which occurs far from detachment, likely remains unaffected.}
{
We also note that Ref.~\cite{huang2023bridging} uses periodic boundary conditions, whereas we use open boundary conditions. However, for the fracture timescale studied here in a small system ($\sim\!\!100$ cells), we do not expect this difference to affect the results, provided that the periodic box is much larger than the tissue size.
}

{
For ease of reproduction and potential future applications, we have wrapped up our code as a \texttt{Python} package \texttt{PyAFV}~\cite{wang2026code}.
Other open-source packages available for simulating vertex/Voronoi-based tissue models include \texttt{cellGPU}~\cite{sussman2017cellgpu} and \texttt{Tyssue}~\cite{theis2021tyssue}, but their standard implementations do not natively support dynamically forming \emph{internal} gaps and detached clusters.
Broader-purpose tissue-simulation frameworks, such as \texttt{Chaste}~\cite{cooper2020chaste}, support multiple agent-based modelling approaches and can represent nonconfluent tissues, but do not provide an implementation of the finite Voronoi model used here and in Refs.~\cite{teomy2018confluent,huang2023bridging}.}

{
Our regularization makes the finite Voronoi model a reliable tool for studying fracture and detachment in nonconfluent tissues, opening many new directions.
For example, in real tissues, the interfacial tensions are not fixed but are dynamically regulated: cell-cell tension is actively remodeled by contractility and adhesion turnover~\cite{cavanaugh2020rhoa}, and cells secrete and degrade extracellular matrix~\cite{perez2024deposited,bonnans2014remodelling} that could feed back on cortical and intercellular tension. Such feedback on $\lambda^{(c)}$ and $\lambda^{(n)}$ could drive confluent-to-nonconfluent transitions and recurrent rupture and healing, an interesting direction for future work.
}

\bigskip
\paragraph*{Code availability.}\hspace{-0.4cm}
The associated purpose-built \texttt{Python} package \texttt{PyAFV} is available on {GitHub} at \href{https://github.com/wwang721/pyafv}{https://github.com/wwang721/pyafv}.
A snapshot of the package and other code required to reproduce this paper have been archived on Zenodo~\cite{wang2026code}.

\begin{acknowledgments}
{Research reported in this publication 
was supported by the National Institute of General Medical Sciences of the National Institutes of Health 
under Award Number R35GM142847. The content is solely the responsibility of the authors and does not 
necessarily represent the official views of the National Institutes of Health.}
This work was carried out at the Advanced Research Computing at Hopkins (ARCH) core facility, which is supported by the National Science Foundation (NSF) Grant No.~OAC1920103. We thank Cody Schimming and Yuntong Zhu for a close reading of the paper and Max Bi for useful conversations. 
\end{acknowledgments}

\appendix
\section{Simulation details}

We simulate the dynamics of both the FV and DP models using the forward Euler (Euler-Maruyama) method~\cite{kloeden1992stochastic}.
Default parameters are given in Table~\ref{tab:param}. All parameters are nondimensionalized using the length scale $L=\sqrt{A_0/\pi}$, energy scale $K_A L^4$, and time scale $1/\mu K_A L^2$.
Note that we are not using the maximum radius $\ell$ as the scaling length unit so that $\ell$ can be set to other values than $1$, such as the optimal $\ell_0$ from minimizing $E_s$ from Eq.~\eqref{eq:single_cell}.

{When simulating the dynamics of the finite Voronoi model, before activating the active motility $v_0$, the initial cell center positions are randomly distributed within a square domain at a packing fraction of about 0.5~\cite{huang2023bridging}. The system is then evolved for $T_\textrm{relax}=20$ time units to reach a steady state. Subsequently, active motility is turned on, and the simulation is run for a duration $T$.}
{The time since relaxation is used in the survival probability and mean survival time calculations.} 

\begin{table}
\caption{\label{tab:param}%
Table of simulation parameters\footnote{These parameters are used throughout the paper; any deviations from them are explicitly specified.}.}
\begin{ruledtabular}
\begin{tabular}{llcc}
Parameter & \qquad Description & Scale & Value\\\hline
$A_0$ & Preferred area & $L^2$ & $\pi$\\
$P_0$ & Preferred perimeter & $L$ & $4.8$\\
$\ell$ & Maximum radius & $L$ & $1$\\
$K_P$ & Perimeter elastic modulus & $K_A L^2$ & $1$\\
$\Lambda$ & Tension difference $\lambda^{(n)}-\lambda^{(c)}$ & $K_A L^3$ & $0.2$\\
$\Delta t$ & Time step & $1/\mu K_A L^2$ & $0.01$\\
$T_\textrm{relax}$ & Initial relaxation time& $1/\mu K_A L^2$ & $20$\\
$T_\mathrm{tot}$ & Total simulation time & $1/\mu K_A L^2$ & $1000$\\
$D_r$ & Rotational noise & $\mu K_A L^2$ & 
$1.33$\\
$N$ & Number of cells & 1 &  $100$\\
\end{tabular}
\end{ruledtabular}
\end{table}

\subsection{Details of finite Voronoi model force calculations}
In the standard/finite Voronoi model, as given in Refs.~\cite{bi2016motility, teomy2018confluent}, an inner vertex connecting three cells $i$, $j$, and $k$ is given by
\begin{equation}
    \mathbf{h}^\textrm{in}=\alpha_i\mathbf{r}_i + \alpha_j\mathbf{r}_j + \alpha_k\mathbf{r}_k.
\end{equation}
The three barycentric coordinates of the circumcenter are
\begin{subequations}
\begin{eqnarray}
    \alpha_i=|\mathbf{r}_j-\mathbf{r}_k|^2(\mathbf{r}_i-\mathbf{r}_j)\cdot(\mathbf{r}_i-\mathbf{r}_k)/D,\\
    \alpha_j=|\mathbf{r}_i-\mathbf{r}_k|^2(\mathbf{r}_j-\mathbf{r}_i)\cdot(\mathbf{r}_j-\mathbf{r}_k)/D,\\
    \alpha_k=|\mathbf{r}_i-\mathbf{r}_j|^2(\mathbf{r}_k-\mathbf{r}_i)\cdot(\mathbf{r}_k-\mathbf{r}_j)/D,
\end{eqnarray}
\end{subequations}
where $D=2|(\mathbf{r}_i-\mathbf{r}_j)\times(\mathbf{r}_j-\mathbf{r}_k)|^2$.
{Fourfold or higher-order Voronoi vertices may occur when four or more cell centers are cocircular, but such configurations are rare numerically and are not relevant to the phenomena studied here.
Our implementation relies on \texttt{SciPy}'s Voronoi construction~\cite{virtanen2020scipy}, which does not robustly identify higher-order vertices under finite numerical precision.
We therefore make an oversimplified assumption that all inner vertices are threefold. In the rare case that an $n$-fold vertex with $n\geqslant4$ is detected, three of the associated cells are selected randomly for the subsequent calculation.}
The derivative of $\mathbf{h}^\textrm{in}$ with respect to $x_i$ is given by
\begin{eqnarray}
    \frac{\partial\mathbf{h}^\textrm{in}}{\partial x_i}&=&\alpha_i\hat{x}+\frac{\partial\alpha_i} {\partial x_i}\mathbf{r}_i+\frac{\partial\alpha_j}{\partial x_i}\mathbf{r}_j+\frac{\partial\alpha_k}{\partial x_i}\mathbf{r}_k \nonumber\\
    &=&\alpha_i\hat{x}+\frac{\partial\alpha_j}{\partial x_i}(\mathbf{r}_j-\mathbf{r}_i)+\frac{\partial\alpha_k}{\partial x_i}(\mathbf{r}_k-\mathbf{r}_i),
\end{eqnarray}
where in the second equation we have used the relation $\alpha_i + \alpha_j+ \alpha_k=1$.
We note this corrects a typo in Ref.~\cite{teomy2018confluent}.
{A similar formula naturally holds for the derivative with respect to $y_i$.} %
Then we need to compute
\begin{eqnarray}
    \frac{\partial \alpha_j}{\partial \mathbf{r}_i} &=& \alpha_j\left[\frac{\mathbf{r}_k-\mathbf{r}_j}{(\mathbf{r}_j-\mathbf{r}_i)\cdot(\mathbf{r}_j-\mathbf{r}_k)} + 2\frac{\mathbf{r}_i-\mathbf{r}_k}{|\mathbf{r}_i-\mathbf{r}_k|^2}\right.\nonumber\\
    &&\left.-2\frac{(\mathbf{r}_j-\mathbf{r}_k)\times\hat{z}}{[(\mathbf{r}_i-\mathbf{r}_j)\times(\mathbf{r}_j-\mathbf{r}_k)]_z}\right].
\end{eqnarray}
The expression for $\partial \alpha_k/\partial \mathbf{r}_i$ follows by exchanging the indices $j$ and $k$.
We have also explicitly shown the expressions for outer vertices in Eqs.~\eqref{eq:h_out} and \eqref{eq:outer_derivative}.

The mechanical force experienced by cell $i$ is given by the gradient of the energy:
\begin{widetext}
\begin{equation}
    f_{i,x}=-\frac{\partial E}{\partial x_i}=-\sum_j\left[2K_A(A_j-A_0)\frac{\partial A_j}{\partial x_i}+2K_P(P_j-P_0)\frac{\partial P_j}{\partial x_i} + \Lambda\frac{\partial P_j^{(n)}}{\partial x_i}\right].
\end{equation}
\end{widetext}
We note that each cell can be decomposed into a polygon (connecting all vertices belonging to that cell), together with additional circular segments ({a segment is defined as the circular sector minus the isosceles triangle formed by the two radii and the chord}), i.e., $A_j=A_j^\textrm{poly}+A_j^\textrm{seg}$ and $P_j=P_j^\textrm{poly}+P_j^{(n)}$.
Thus, the derivatives of the area and perimeter can be decomposed into two corresponding parts:
\begin{equation}
    \frac{\partial A_j}{\partial x_i}=\frac{\partial A_j^\textrm{poly}}{\partial x_i}+\frac{\partial A_j^\textrm{seg}}{\partial x_i},~~~\frac{\partial P_j}{\partial x_i}=\frac{\partial P_j^\textrm{poly}}{\partial x_i}+\frac{\partial P_j^{(n)}}{\partial x_i}. 
\end{equation}
For the polygon part,
\begin{equation}
    \frac{\partial A_j^\textrm{poly}}{\partial x_i} = \sum_m\frac{\partial A_j^\textrm{poly}}{\partial \mathbf{h}_m}\cdot\frac{\partial \mathbf{h}_m}{\partial x_i},~~~\frac{\partial P_j^\textrm{poly}}{\partial x_i} = \sum_m\frac{\partial P_j^\textrm{poly}}{\partial \mathbf{h}_m}\cdot\frac{\partial \mathbf{h}_m}{\partial x_i},
\end{equation}
where the sum runs over all vertices (both inner and outer), as in Eq.~\eqref{eq:fx}. {The derivatives $\partial\mathbf{h}_m/\partial x_i$ are already given above, while the derivatives ${\partial A_j^\textrm{poly}}/{\partial \mathbf{h}_m}$ and ${\partial P_j^\textrm{poly}}/{\partial \mathbf{h}_m}$ are provided in Eq.~(B15) of Ref.~\cite{teomy2018confluent}.}

We now compute the contribution from the segment part. For a cell $j$ with a total segment area $A_j^\textrm{seg}$ and total arc length $P_j^{(n)}$, consider a segment between two neighboring outer vertices $\mathbf{h}^\textrm{out}_m$ and $\mathbf{h}^\textrm{out}_{m+1}$ (ordered clockwise). Let $\phi_m^{j}$ denote the angle between the vector $\mathbf{u}=\mathbf{h}^\textrm{out}_m - \mathbf{r}_j$ and the $x$-axis, i.e., $\phi_m^j=\operatorname{atan2}(u_y, u_x)\in(-\pi,\pi]$. The arc length between the two vertices is then $\Delta P=\ell\Delta\phi$, and the area of the corresponding segment is
\begin{equation}
    \Delta A = \frac{\ell^2}{2}(\Delta\phi-\sin\Delta\phi),
\end{equation}
where $\Delta\phi\equiv \phi_m^{j}-\phi_{m+1}^{j}\pmod{2\pi}$. Thus we have
\begin{equation}
    \frac{\partial P_j^{(n)}}{\partial\phi_m^{j}}=\ell,~~~\frac{\partial A^\textrm{seg}_j}{\partial\phi_{m}^{j}}=\frac{\ell^2}{2}(1-\cos\Delta\phi),
\end{equation}
and ${\partial P_j^{(n)}}/{\partial\phi_{m+1}^{j}} = -{\partial P_j^{(n)}}/{\partial\phi_{m}^{j}}$, ${\partial A^\textrm{seg}_j}/{\partial\phi_{m+1}^{j}}=-{\partial A^\textrm{seg}_j}/{\partial\phi_{m}^{j}}$.
Because an outer vertex such as $\mathbf{h}_m^\textrm{out}$ is shared by cell $j$ and a neighboring cell $i$ [see Eq.~\eqref{eq:h_out}], the angle $\phi_m^{j}$ depends on both $\mathbf{r}_i$ and $\mathbf{r}_j$. With the definition $\mathbf{u}=\mathbf{h}_m^\textrm{out}-\mathbf{r}_j=|\mathbf{u}|(\cos\phi_m^j,\sin\phi_m^j)$ and $|\mathbf{u}|=\ell$, we have
\begin{equation}\label{eq:du_dx}
    \frac{\partial \mathbf{u}}{\partial x_i}=\frac{\partial \mathbf{h}_m^\textrm{out}}{\partial x_i}=\ell(-\sin\phi_m^j,\cos\phi_m^j)\frac{\partial\phi_m^j}{\partial x_i},
\end{equation}
where $(-\sin\phi_m^j,\cos\phi_m^j)=-(\mathbf{u}\times\hat{z})/|\mathbf{u}|$ is a unit vector perpendicular to $\mathbf{u}$.
Thus, taking the dot product of Eq.~\eqref{eq:du_dx} with $(-\sin\phi_m^j,\cos\phi_m^j)/\ell$, we obtain the derivative of $\phi_m^j$ with respect to $x_i$:
\begin{equation}
    \frac{\partial\phi_m^{j}}{\partial x_i}=-\frac{1}{\ell^2}[(\mathbf{h}_m^\textrm{out}-\mathbf{r}_j)\times\hat{z}]\cdot\frac{\partial\mathbf{h}_m^\textrm{out}}{\partial x_i},
\end{equation}
and the $y$-component has an analogous form.
Since $\mathbf{u}$ (and hence $\phi^j_m$) depends only on the relative position $\mathbf{r}_i-\mathbf{r}_j$, we also have $\partial{\phi^{j}_m}/\partial \mathbf{r}_j=-\partial{\phi^{j}_m}/\partial \mathbf{r}_i$. With all these derivatives in hand, we can then compute the segment contributions to the derivative of the area and perimeter:
\begin{eqnarray}
    \frac{\partial A_j^\textrm{seg}}{\partial x_i}=\sum_{\mathbf{h}_m^\textrm{out}}\frac{\partial A_j^\textrm{seg}}{\partial\phi_m^{j}}\frac{\partial\phi^{j}_m}{\partial x_i},~\frac{\partial P_j^{(n)}}{\partial x_i}=\sum_{\mathbf{h}^\textrm{out}_m}\frac{\partial P_j^{(n)}}{\partial \phi_m^j}\frac{\partial \phi_m^j}{\partial x_i}.
\end{eqnarray}
We have verified the numerical force calculations using \texttt{PyAFV}~\cite{wang2026code} for given configurations $\{\mathbf{r}_i\}$ against the \texttt{MATLAB} code from Ref.~\cite{huang2023bridging}.

\subsection{Details of deformable polygon model}\label{app:DP}
In our simulations of the deformable polygon model, each cell is represented by $M=100$ vertices, {which we found to be sufficiently large that the detachment forces are not changed relevantly by refining the number of vertices.}  Enumerating the vertices $\{\mathbf{h}_m\}$ of cell $i$ in clockwise order, the force exerted on vertex $m$ is given by
\begin{equation}
    \mathbf{f}_m=-\frac{\partial E}{\partial \mathbf{h}_m}.
\end{equation}
We need the derivative of area $A_i$ with respect to $\mathbf{h}_m$~\cite{teomy2018confluent}: 
\begin{equation}
    \frac{\partial A_i}{\partial\mathbf{h}_m}=\frac{(\mathbf{h}_{m-1}-\mathbf{h}_{m+1})\times\hat{z}}{2},
\end{equation}
and derivatives of $P_m$ and $P_{m-1}$ with respect to $\mathbf{h}_m$ \cite{teomy2018confluent}:
\begin{equation}
    \frac{\partial P_m}{\partial \mathbf{h}_m}=\frac{\mathbf{h}_m-\mathbf{h}_{m+1}}{|\mathbf{h}_m-\mathbf{h}_{m+1}|},~~~\frac{\partial P_{m-1}}{\partial \mathbf{h}_m}=\frac{\mathbf{h}_m-\mathbf{h}_{m-1}}{|\mathbf{h}_m-\mathbf{h}_{m-1}|},
\end{equation}
where $P_{m}=|\mathbf{h}_m-\mathbf{h}_{m+1}|$, and $\mathbf{h}_{m-1}$ and $\mathbf{h}_{m+1}$ are two adjacent vertices of $\mathbf{h}_m$ in cell $i$.
External forces are applied quasi-statically, starting from zero, and the system is evolved for $5\times10^4$ steps (step size $\Delta t=0.001$) to reach a sequence of steady states.
Vertices are resampled every $1000$ steps by redistributing them along the free-boundary polyline to achieve approximately uniform spacing between adjacent vertices~\cite{weng2022convergent}.
We record the force at which the contact length $P^{(c)}$ falls below the typical segment length between vertices, $\ell_c =2\pi\ell_0/M$, as the detachment force $f_\textrm{detach}^\textrm{DP}$.

\section{Analytical details}\label{app:analyt}
{
To derive the interaction force $f=-\partial E_s/\partial \epsilon$ of a cell doublet [Eq.~\eqref{eq:force}], we first compute the necessary derivatives. For the angle spanning the non-contacting region $\phi=2\pi-2\operatorname{atan2}\bigl(\sqrt{\ell^2-(\ell-\epsilon)^2},\ell-\epsilon\bigr)$, we obtain
\begin{equation}
    \frac{\partial\phi}{\partial\epsilon}=-\frac{2}{\sqrt{(2\ell-\epsilon)\epsilon}}.
\end{equation}
This gives
\begin{equation}
    \frac{\partial A}{\partial \epsilon} = -2\sqrt{(2\ell-\epsilon)\epsilon},~~~\frac{\partial P}{\partial \epsilon}=-\frac{2\epsilon}{\sqrt{(2\ell-\epsilon)\epsilon}}.
\end{equation}
We therefore see that the divergent contribution arises from the non-contacting perimeter $P^{(n)}=\ell\phi$ rather than from the area $A$ or the total perimeter $P$.}

{We can further take the limit $\epsilon\to0^+$ in Eq.~\eqref{eq:force}, yielding the asymptotic form
\begin{equation}\label{eq:force_0}
    2\sqrt{2\ell\epsilon}\left[2K_A(\pi\ell^2-A_0) + K_P\frac{(2\pi\ell-P_0)}{\ell}\right] + {\Lambda}\sqrt{\frac{2\ell}{\epsilon}},
\end{equation}
which diverges for $\Lambda\neq 0$, as shown in Fig.~\ref{fig:doublet}b.
This implies that, with $\Lambda\neq 0$, contacts become increasingly hard to break as their length shrinks to zero in the AFV model.
Note that by using $K_A$ and $L=\ell$ to nondimensionalize terms in the square bracket, we recover the dimensionless criterion for attractive forces given in Ref.~\cite{teomy2018confluent} for the $\Lambda=0$ case:
$
    2(\pi-\tilde{A}_0) + \tilde{K}_P(2\pi-\tilde{P}_0)>0 \label{eq:asymptotic_appendix}
$.}

{
\section{Additional figures}
}

\begin{figure}[htbp]
\includegraphics[width=0.48\textwidth]{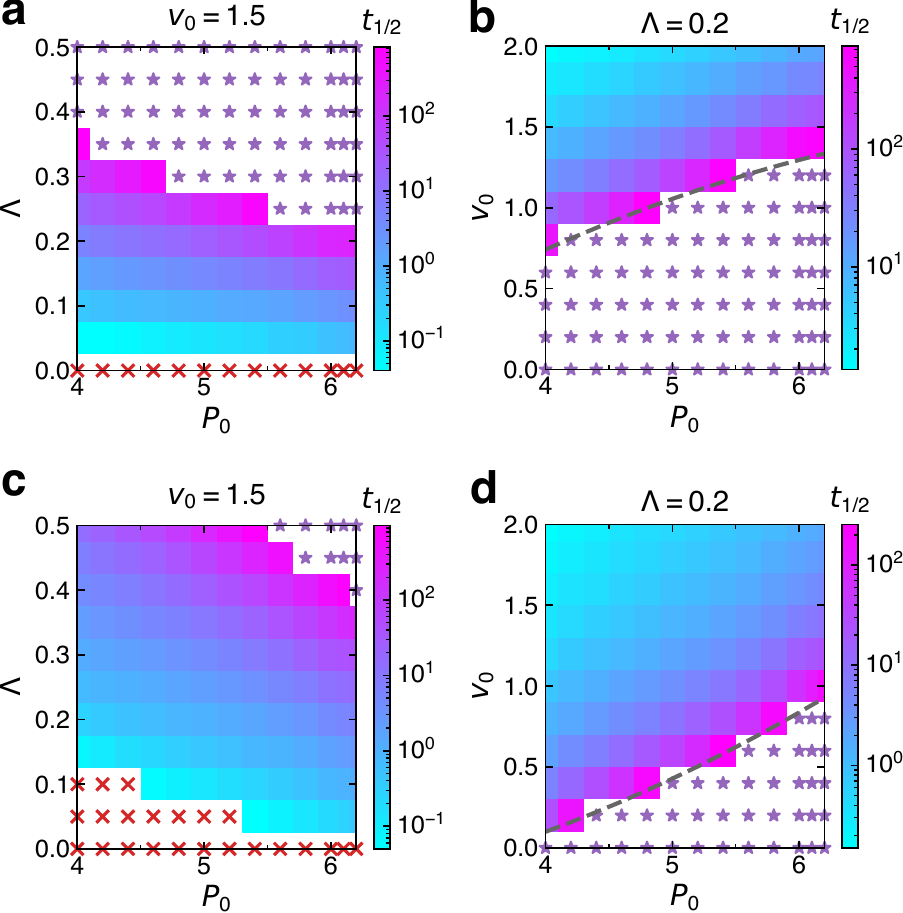}
\caption{{\textbf{Dependence of the fracture timescale on the cutoff threshold $\delta$ under Strategy 1. a,b} are phase diagrams of $t_{1/2}$ for $\delta=0.3$.
\textbf{c,d} correspond to the same phase diagrams for $\delta=0.6$.
Default parameters: $K_P=1$, $A_0=\pi, D_r=1.33$.}
}
\label{fig:SM_vary_delta}
\end{figure}

\begin{figure}[htbp]
\includegraphics[width=0.45\textwidth]{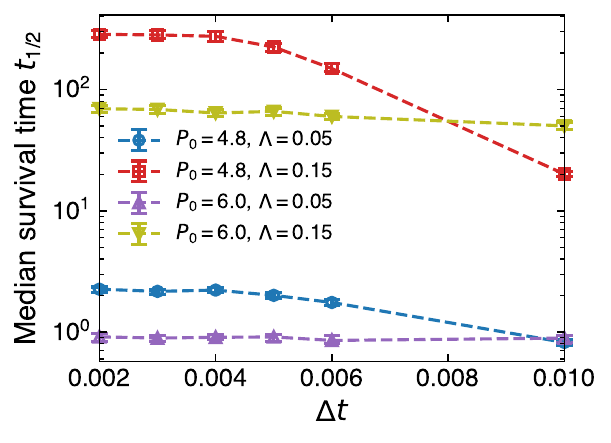}
\caption{{\textbf{Convergence of the fracture timescale as $\Delta t\to 0$ under Strategy 2.} The values of $t_{1/2}$ at four selected points in the $P_0$--$\Lambda$ phase diagrams converge as the simulation time step $\Delta t$ approaches zero.
Default parameters: $K_P=1$, $A_0=\pi, D_r=1.33, v_0=1.5$.}
}
\label{fig:SM_vary_dt}
\end{figure}

\bibliography{refs}%

\end{document}